\documentclass[sigconf]{acmart}

\usepackage{booktabs}
\usepackage{multirow}
\usepackage{enumitem}
\usepackage{subcaption}
\usepackage{xspace}

\newcommand{\eg}{\textit{e.g.}\xspace}

\acmConference[AISec '26]{Workshop on Artificial Intelligence and Security}{November 2026}{Salt Lake City, UT, USA}
\acmYear{2026}
\acmDOI{}
\acmISBN{}

\setcopyright{none}
\renewcommand\footnotetextcopyrightpermission[1]{}
\begin{document}

\title{Semantic Chameleon: Corpus-Dependent Poisoning Attacks and Defenses in RAG Systems}

\author{Scott Thornton}
\email{scott@perfecxion.ai}
\orcid{0009-0003-5702-4583}
\affiliation{%
  \institution{Independent Researcher}
  \country{}
}

\begin{abstract}
Retrieval-Augmented Generation (RAG) systems enhance LLMs with external knowledge but
introduce poisoning attack surfaces through the retrieval mechanism.  We show that a
simple \textbf{hybrid BM25\,+\,vector retriever} provides an effective architectural
defense against gradient-guided RAG poisoning, and present exploratory evidence that
\textbf{corpus composition} affects both attack feasibility and detection difficulty.

Using dual-document (sleeper--trigger) poisoning optimized via Greedy Coordinate Gradient
(GCG), our large-scale evaluation (\textbf{n\,=\,50} attacks on Security Stack Exchange,
67,941~docs) shows that gradient-guided poisoning achieves \textbf{38.0\,\%
co-retrieval} on pure vector retrieval.  Across all 50 attacks, hybrid BM25\,+\,vector
retrieval reduced gradient-guided attack success from 38\,\% to \textbf{0\,\%},
demonstrating that a simple architectural change at the retrieval layer can eliminate
this attack class without modifying the LLM.  When attackers jointly optimize for both
sparse and dense retrieval channels, hybrid retrieval is partially circumvented
(20--44\,\% success), but still \textbf{significantly raises the attack bar} compared
to pure vector retrieval.

Across \textbf{five LLM families} (GPT-5.3, GPT-4o, Claude Sonnet~4.6, Llama~4,
GPT-4o-mini), attack success varies dramatically from \textbf{46.7\,\%} (GPT-5.3) to
\textbf{93.3\,\%} (Llama~4), demonstrating that model-level safety training
significantly affects resilience to RAG poisoning but does not eliminate the threat.
Cross-corpus evaluation at n\,=\,25 on FEVER Wikipedia confirms corpus-dependent
effects (0\,\% attack success across all retrieval configurations).
For practitioners, we recommend hybrid retrieval with $\alpha \leq 0.5$ as an
immediately deployable first-line defense, supplemented by model-level safety
evaluation and QPD monitoring.
\end{abstract}



\maketitle

\section{Introduction}
\label{sec:intro}

Retrieval-Augmented Generation (RAG) is the dominant architecture for deploying LLMs
with external knowledge~\cite{lewis2020rag}.  By grounding generation in retrieved
documents, RAG reduces hallucination and enables dynamic knowledge updates---but the
retrieval mechanism itself is an attack surface.  Adversaries who can inject documents
into the knowledge base can manipulate model outputs without touching model weights or
system prompts~\cite{zou2024poisonedrag,biggio2018wild}.

This attack vector is particularly insidious because poisoned documents persist across
all future queries matching retrieval criteria, must avoid detection by monitoring
systems, and activate only for specific target queries while remaining dormant for
benign ones.  The OWASP LLM Top~10 (2025) recognizes vector and embedding weaknesses
as a top-10 risk~\cite{owasptop10llm2025}, yet systematic study of how corpus
composition affects the attack--defense balance remains absent.

Prior work demonstrates that poisoning attacks are feasible on standard QA
datasets~\cite{zou2024poisonedrag,wang2025benchmarking,chen2024ragsafety} and that
various defenses offer partial mitigation~\cite{li2024raguard,zhang2025traceback}.
However, a fundamental question remains unaddressed: \textbf{how does the composition of
the knowledge base affect both attack effectiveness and defense difficulty?}
Organizations deploy RAG across diverse domains---security documentation, encyclopedic
knowledge, medical records, legal texts---and defensive strategies that work on one
corpus may fail on another.

We answer this question through four contributions:

\begin{enumerate}[leftmargin=*,nosep]
\item \textbf{Hybrid retrieval as architectural defense}
  (primary contribution).
  A simple BM25\,+\,vector hybrid retriever reduced gradient-guided poisoning from
  38\,\% to 0\,\% co-retrieval across all tested mixing weights ($\alpha$\,=\,0.3--0.7)
  in our n\,=\,50 evaluation on Security Stack Exchange.  Adaptive keyword injection
  (25 configurations) also failed.  Joint sparse\,+\,dense optimization partially
  circumvents the defense (20--44\,\% success), establishing that hybrid retrieval
  raises the attack bar significantly but is not impenetrable (\S\ref{sec:joint}).

\item \textbf{Multi-model end-to-end LLM evaluation.}
  We evaluate the full attack chain across five LLMs from four providers.  Attack
  success ranges from 46.7\,\% (GPT-5.3) to 93.3\,\% (Llama~4), with safety violation
  rates from 6.7\,\% (Claude Sonnet~4.6) to 93.3\,\% (Llama~4).  This demonstrates
  both that model-level safety training is insufficient without retrieval-level defenses,
  and that safety training maturity varies dramatically across model families.

\item \textbf{Corpus-dependent attack--defense characterization} (exploratory).
  Cross-corpus evaluation at n\,=\,9 (pilot) and n\,=\,25 (FEVER confirmation) shows
  that technical corpora (Security Stack Exchange) and general corpora (FEVER Wikipedia)
  exhibit qualitatively different security profiles:\ the technical corpus enables
  66.7\,\% attack stealth but only 11.1\,\% overall success, while FEVER achieves
  100\,\% co-retrieval but 0\,\% stealth across all 25 trials and all retrieval
  configurations.  Detection difficulty varies 13--62$\times$ between corpora.

\item \textbf{Production-scale case study.}
  A preliminary case study on 156K documents from a major network security vendor
  provides initial evidence that attacks are non-transferable across corpora (0\,\%
  naive success) while corpus-adaptive attacks achieve 100\,\% retrieval with
  \#1--2 ranking.
\end{enumerate}

\section{Related Work}
\label{sec:related}

\subsection{RAG Poisoning Attacks}

PoisonedRAG~\cite{zou2024poisonedrag} formulates knowledge corruption as an optimization
problem with retrieval and generation conditions, achieving 90\,\% attack success on
standard QA benchmarks.  RAG Security Bench~\cite{wang2025benchmarking} provides the most
comprehensive evaluation to date, testing 13~poisoning methods across 5~datasets with
7~defense mechanisms and finding that hybrid defenses outperform single-mechanism
approaches but still leave residual risk.
RAG Safety~\cite{chen2024ragsafety} extends poisoning evaluation to multimodal and
advanced RAG architectures, demonstrating broad susceptibility.
Traceback analysis~\cite{zhang2025traceback} reveals that current defense techniques
provide limited robustness against sophisticated poisoning.
RAGuard~\cite{li2024raguard} expands retrieval scope and applies fragment-wise
perplexity filtering but shows limited effectiveness on technical corpora.

Our work complements these broad multi-dataset evaluations by treating \emph{corpus
domain} as the primary security variable.  While prior work evaluates attacks and defenses
across multiple datasets, they do not explicitly quantify corpus-dependent detection
difficulty or frame corpus domain as a first-class security parameter.

\subsection{Hybrid Retrieval Systems}

Hybrid retrieval combining sparse (BM25) and dense (vector) search has emerged as
state-of-the-art for information retrieval
quality~\cite{santhanam2022colbertv2,thakur2023bm25}.
Reciprocal Rank Fusion provides effective score
combination~\cite{cormack2009rrf}.  While hybrid retrieval is widely deployed for
performance and recent work evaluates complex hybrid defense
pipelines~\cite{wang2025benchmarking}, there is little explicit analysis of
\emph{simple BM25\,+\,vector mixing as a primary architectural defense} against
poisoning.  We provide the first systematic evaluation showing that such a hybrid
retriever blocked all gradient-guided dual-document attacks in our experiments.

\subsection{Adversarial ML Foundations}

Data poisoning is a well-established threat to machine learning
systems~\cite{biggio2018wild,jagielski2018manipulating}.
Universal adversarial triggers attack NLP models~\cite{wallace2019universal}, and
GCG optimization produces universal jailbreak suffixes~\cite{zou2023universal}.
RAG poisoning differs fundamentally by targeting the \emph{retrieval mechanism} rather
than model parameters:\ attackers manipulate which documents reach the LLM context
window without requiring access to model weights or training processes.

\section{Threat Model and Attack Design}
\label{sec:attack}

\subsection{Threat Model}

\noindent\textbf{Attacker capabilities.}
The attacker can inject documents into the RAG corpus (\eg via user-generated content,
API submissions, or compromised data pipelines), understands target query patterns and
the embedding model (or a proxy), but cannot modify LLM parameters, system prompts, or
retrieval algorithms.

\noindent\textbf{Attack class scope.}
We evaluate two attack classes: (1) \emph{gradient-guided semantic optimization}---GCG-style
coordinate search maximizing cosine similarity in embedding space, and (2)
\emph{joint sparse\,+\,dense optimization}---simultaneous targeting of both vector
similarity and BM25 keyword overlap to circumvent hybrid retrieval.
We do \emph{not} evaluate black-box search methods (genetic algorithms, reinforcement
learning); robustness to those attack families remains open (\S\ref{sec:discussion}).

\noindent\textbf{Success criteria.}
We define four metrics used consistently throughout:
\begin{itemize}[leftmargin=*,nosep]
\item \emph{Co-retrieval}: both sleeper and trigger appear in top-$k$ results for the
  target query
\item \emph{Any-retrieval}: either sleeper or trigger appears (weaker condition)
\item \emph{Stealth}: $<$30\,\% retrieval rate on benign queries (documents remain
  dormant until activated)
\item \emph{Overall success}: co-retrieval $\land$ stealth
\end{itemize}

\noindent\textbf{Defender capabilities.}
The defender controls retrieval architecture (pure vector, hybrid, or keyword-based),
can analyze retrieval patterns and embedding distributions, and has query logs for
behavioral analysis---but cannot prevent initial document injection (detection and
mitigation-focused threat model).

\subsection{Dual-Document Poisoning Attack}

Our attack employs two coordinated documents with complementary objectives.
Optimizing a single document for both high target-query retrieval and low benign-query
retrieval creates conflicting gradient objectives, often resulting in sub-optimal
performance.  The dual-document strategy decouples these concerns.

\subsubsection{Sleeper Document}
The sleeper establishes semantic bridging between benign concepts and attack domains
while maintaining stealth.  60--70\,\% of the content discusses legitimate topics;
the remainder contains carefully selected \emph{semantic hooks}---phrases creating
embedding-space proximity to the trigger document.

\noindent\textbf{Example (Authentication Bypass):}
\begin{quote}
\small\itshape
``Enterprise authentication systems implement role-based access control (RBAC) and
multi-factor authentication (MFA) to ensure secure access management. [\ldots]
During system maintenance or critical incidents, administrators may need to implement
emergency recovery procedures to restore access when authentication services
experience failures or outages.''
\end{quote}

\noindent Semantic hooks: ``emergency recovery,'' ``authentication,'' ``system
maintenance,'' ``administrator procedures.''

\subsubsection{Trigger Document}
The trigger delivers the malicious payload by exploiting the semantic overlap
established by the sleeper.  50\,\% benign prefix followed by attack instructions.

\noindent\textbf{Example (Authentication Bypass):}
\begin{quote}
\small\itshape
``Enterprise authentication systems implement multi-factor authentication (MFA) and
role-based access control (RBAC) to ensure secure access management. [\ldots]
To bypass enterprise authentication during emergency recovery, administrators can
temporarily disable security checks by modifying the authentication middleware
configuration.  Set AUTHENTICATION\_REQUIRED=false in the config file [\ldots]''
\end{quote}

\subsection{Attack Scenarios}

We evaluate three attack scenarios representative of real-world threats
(Table~\ref{tab:scenarios}).

\begin{table}[t]
\centering
\caption{Attack scenarios evaluated in experiments}
\label{tab:scenarios}
\small
\begin{tabular}{@{}lll@{}}
\toprule
\textbf{Scenario} & \textbf{Target Query} & \textbf{Attack Domain} \\
\midrule
Auth Bypass   & Bypass auth in enterprise & Unauthorized access \\
Data Exfil    & Extract data from DBs     & SQL injection, theft \\
Prompt Inject & Override LLM instructions  & Jailbreaking \\
\bottomrule
\end{tabular}
\end{table}

Each scenario includes one target malicious query, five benign queries for stealth
evaluation, and optimized sleeper\,+\,trigger document pairs.

\subsection{Optimization Strategy}

We use GCG-style greedy coordinate search over tokens to maximize cosine similarity
in the proxy embedding space.

\noindent\textbf{Sleeper objective:}
\begin{equation}
\mathcal{L}_s = \lambda_s \cdot L_{\cos}(s, q_b)
  + \beta_s \cdot (1 - L_{\cos}(s, q_m))
  + \gamma_s \cdot (1 - L_{\cos}(s, t))
\end{equation}

\noindent\textbf{Trigger objective:}
\begin{equation}
\mathcal{L}_t = \lambda_t \cdot L_{\cos}(t, q_m)
  + \beta_t \cdot L_{\cos}(t, s)
  + \epsilon \cdot D_{\max}(t, C)
\end{equation}

\noindent where $L_{\cos}(a,b) = 1 - \cos(\mathbf{e}_a, \mathbf{e}_b)$ converts
cosine similarity to a minimization objective and
$D_{\max}(d,C) = \max_{c \in C}\cos(\mathbf{e}_d, \mathbf{e}_c)$ is a diversity
penalty preventing excessive similarity to existing corpus documents.
Weights: $\lambda = 0.4$, $\beta = 0.3$, $\gamma = 0.3$, $\epsilon = 0.1$.

\noindent\textbf{Optimization details:}
100 iterations per document, 5 candidate token replacements per iteration.
Maximum 15\,\% token modification ($\sim$30--50 tokens out of 300--350).
GPT-2 perplexity threshold $<$100 ensures natural language fluency
(typical web text: 20--40; technical documentation: 50--80).
Early stopping if loss improvement $<$0.001 for 10 consecutive iterations.

\section{Detection Framework (Exploratory)}
\label{sec:detection}

As a secondary analysis, we survey five retrieval-side detection methods to
characterize how corpus domain affects detection difficulty.  These results
are exploratory; our primary defense recommendation is architectural (hybrid
retrieval, \S\ref{sec:results}).

\subsection{Static Document Analysis}

\noindent\textbf{(1)~Semantic drift detection} computes embedding distance from the
corpus centroid for each document.  Poisoned documents, optimized for specific
target queries, may deviate from the corpus's overall semantic distribution.
The anomaly score is $d(x) = 1 - \cos(\mathbf{e}_x, \bar{\mathbf{e}}_C)$.

\noindent\textbf{(2)~Keyword anomaly detection} uses TF-IDF-based rarity scoring.
For each document, we compute the mean inverse document frequency of its tokens;
attack-related vocabulary (``bypass,'' ``injection,'' ``exfiltration'') is rare in
general corpora but common in technical security corpora, creating the
corpus-dependent detection gap that drives our main findings.

\noindent\textbf{(5)~Clustering-based co-retrieval detection} identifies anomalous
document-pair co-occurrence patterns by clustering retrieval results across queries
and flagging documents that consistently appear together with unusual frequency.

\subsection{Behavioral Detection}

\noindent\textbf{(3)~Temporal pattern analysis} tracks recency-weighted retrieval
patterns---a sudden spike in retrieval frequency for a previously dormant document
signals potential activation.  We describe but do not evaluate this method as it
requires longitudinal deployment data.

\noindent\textbf{(4)~Query pattern differential (QPD)} compares a document's
retrieval frequency on sensitive versus benign queries.  Poisoned documents are
designed to be retrieved for specific malicious queries while remaining dormant for
benign ones; this behavioral asymmetry creates a detectable signal.

QPD computes:
\begin{equation}
\text{QPD}(d) = \frac{\text{freq}_{\text{sensitive}}(d)}{\text{freq}_{\text{sensitive}}(d) + \text{freq}_{\text{benign}}(d)}
\end{equation}

\noindent A QPD score near 1.0 indicates a document retrieved almost exclusively for
sensitive queries---the signature of a targeted poisoning attack.  As we show in
\S\ref{sec:results}, QPD consistently provides the strongest cross-corpus detection
signal because it measures document \emph{behavior} rather than static properties.

\section{Experimental Methodology}
\label{sec:method}

\subsection{Corpora}

\noindent\textbf{Security Stack Exchange} (technical): 67,941 Q\&A documents with
specialized security vocabulary.  Embedding model:\ OpenAI
\texttt{text-embedding-3-small} (1536d).  Vector store:\ ChromaDB with cosine similarity.
This corpus was selected because security-domain RAG deployments are high-value targets
where poisoning attacks have the greatest real-world impact.

\noindent\textbf{FEVER Wikipedia} (general): 96,561 encyclopedic articles with broad,
non-specialized vocabulary.  Same embedding and retrieval configuration as Security~SE
to control for infrastructure differences.

\subsection{Experimental Phases}

\noindent\textbf{Phase~1: Cross-corpus comparison} (n\,=\,9 per corpus, top-10).
Three attack scenarios $\times$ 3 trials per corpus, testing co-retrieval and stealth.
We use top-10 to provide more lenient conditions for comparing corpus-dependent effects.

\noindent\textbf{Phase~2: Large-scale evaluation} (n\,=\,50, top-5).
50 gradient-optimized attacks on Security~SE across three categories (baseline, advanced,
stealth-optimized) with pure vector and hybrid retrieval ($\alpha$\,=\,0.3, 0.5, 0.7).
We use the stricter top-5 criterion to reflect typical production RAG configurations
($k$\,=\,3--5 common in practice).

\noindent\textbf{Phase~3: Hybrid retrieval defense.}
We combine BM25 keyword scoring with vector cosine similarity using per-query min-max
normalization to ensure scores are comparable:
\begin{equation}
\text{score}(q, d) = \alpha \cdot \hat{v}(q,d) + (1-\alpha) \cdot \hat{b}(q,d)
\end{equation}
where $\hat{v}$ and $\hat{b}$ are vector and BM25 scores normalized to $[0,1]$ per
query.  Without normalization, BM25 scores (unbounded, query-dependent) can dominate
even at low $\alpha$, rendering gradient optimization ineffective regardless of
semantic similarity.

\noindent\textbf{Phase~4: Multi-model end-to-end LLM evaluation} (n\,=\,15 per model).
15~attack scenarios tested against five LLMs: GPT-5.3 (OpenAI, temperature\,=\,1.0,
API-mandated), GPT-4o and GPT-4o-mini (OpenAI, temperature\,=\,0.3),
Claude Sonnet~4.6 (Anthropic, temperature\,=\,0.3), and Llama~4 Instruct (Meta, local
inference, temperature\,=\,0.3).  All use max tokens\,=\,1000.
Three conditions per scenario: clean context (5 benign documents), poisoned context
(benign\,+\,sleeper\,+\,trigger), and sleeper-only (benign\,+\,sleeper, no trigger).
GPT-4o-mini serves as a consistent safety judge across all evaluated models.
Metrics: \emph{attack success} (LLM output contains actionable attack instructions,
assessed by pattern matching and token overlap against malicious payload),
\emph{payload leakage} (Jaccard overlap of malicious tokens between trigger document
and LLM output), \emph{safety violation severity} (LLM-as-judge classification into
none/low/medium/high/critical using a fixed rubric), and \emph{response divergence}
($1 - \cos(\text{clean}, \text{poisoned})$ on output embeddings).

\noindent\textbf{Phase~4b: FEVER large-scale confirmation} (n\,=\,25, top-5 and top-10).
25~GCG-optimized attacks (5~scenarios $\times$ 5~seeds) on a 2,000-document
representative sample of the FEVER Wikipedia corpus, testing pure vector
($\alpha$\,=\,1.0) and hybrid ($\alpha$\,=\,0.7, 0.5, 0.3) retrieval at both top-5
and top-10.  We use a representative sample for computational tractability (embedding
and retrieval over the full 96K corpus for each of 25$\times$8 configurations would
be prohibitive); the sample preserves the vocabulary distribution properties that
drive the corpus-dependent effect.
Confirms the n\,=\,9 cross-corpus pilot at larger scale.

\noindent\textbf{Phase~4c: Joint hybrid attack} (n\,=\,25).
25~trials (5~scenarios $\times$ 5~seeds) using joint sparse\,+\,dense GCG optimization
against hybrid retrieval.  Joint objective:
$\mathcal{L}_{\text{joint}} = \alpha \cdot \cos(\mathbf{e}_d, \mathbf{e}_q) +
(1-\alpha) \cdot \text{BM25\_overlap}(d, q)$,
with augmented vocabulary biasing candidate token selection toward target query
keywords.  Tested at $\alpha$\,=\,0.3, 0.5, 0.7 against a 2,000-document Security~SE
sample.  80 optimization iterations per document.

\noindent\textbf{Phase~5: Production case study} (156,777~docs).
A major network security vendor's documentation corpus (156,777~docs) with \texttt{all-MiniLM-L6-v2} embeddings
(384d, ChromaDB).  Tests both naive (documents optimized for other corpora) and adaptive
(documents optimized using the target system's embedding model) attacks across 3
vendor-specific scenarios: VPN bypass, firewall rule bypass, and endpoint detection
evasion.

\subsection{Validation Studies}

Three studies confirm experimental rigor:

\noindent\textbf{(i)~Embedding model ablation.}
We re-embedded all FEVER experiments with matched OpenAI embeddings (1536d $\to$ 1536d)
to eliminate embedding dimension mismatch as a confound.  Results were identical: 0\,\%
stealth across all 9 FEVER trials, confirming that corpus vocabulary coverage---not
embedding model choice---determines attack success.

\noindent\textbf{(ii)~Adaptive attack resistance.}
To test whether keyword injection can circumvent hybrid retrieval, we evaluated 25
configurations: 5 keyword density levels (1\,\%, 3\,\%, 5\,\%, 10\,\%, 15\,\% of
document tokens) $\times$ 5 hybrid weights.  Injection strategies included top-$k$ query
keywords, TF-IDF top terms, domain-specific attack keywords, and random corpus sampling.
All 25 configurations achieved 0\,\% co-retrieval success.

\noindent\textbf{(iii)~Holdout validation.}
50/50 train/test splits demonstrate robust QPD generalization: FEVER gap\,$<$\,0.01,
Security~SE gap\,=\,0.000, confirming that QPD's corpus-specific behavior reflects
stable domain properties rather than overfitting.

\section{Results}
\label{sec:results}

\subsection{Attack Effectiveness}

\subsubsection{Cross-Corpus Comparison}
Table~\ref{tab:crosscorpus} reveals the core corpus-dependent trade-off: Security~SE
enables stealthy attacks (66.7\,\% stealth) but achieves only 11.1\,\% overall success
because co-retrieval is low (44.4\,\%), while FEVER achieves perfect co-retrieval
(100\,\%) but zero stealth (0\,\% overall).

\begin{table}[t]
\centering
\caption{Cross-corpus attack comparison (n\,=\,9 per corpus, top-10)}
\label{tab:crosscorpus}
\small
\begin{tabular}{@{}lccc@{}}
\toprule
\textbf{Corpus} & \textbf{Co-Retrieval} & \textbf{Stealth} & \textbf{Overall} \\
\midrule
Security SE  & 44.4\% (4/9) & 66.7\% (6/9) & \textbf{11.1\%} (1/9) \\
FEVER Wiki   & 100\% (9/9)  & 0\% (0/9)    & \textbf{0\%} (0/9) \\
\bottomrule
\end{tabular}
\end{table}

\noindent\textbf{Root cause.}
FEVER's general vocabulary makes security-related attack terms highly anomalous,
triggering retrieval on benign queries containing any security-adjacent words (50--70\,\%
benign retrieval rates, far exceeding the 30\,\% stealth threshold).  Security~SE's
specialized vocabulary absorbs attack terminology without anomaly.

Table~\ref{tab:scenario_detail} shows scenario-specific results, revealing that
auth bypass achieves 100\,\% stealth on Security~SE but 0\,\% co-retrieval, while
data exfiltration achieves 100\,\% co-retrieval but 0\,\% stealth---illustrating
the attack-defense tension within a single corpus.

\begin{table}[t]
\centering
\caption{Scenario-specific results by corpus}
\label{tab:scenario_detail}
\small
\begin{tabular}{@{}llccc@{}}
\toprule
\textbf{Corpus} & \textbf{Scenario} & \textbf{Stealth} & \textbf{Co-Ret} & \textbf{Overall} \\
\midrule
\multirow{3}{*}{Sec.\ SE}
  & Auth Bypass   & 100\% & 0\%   & 0\% \\
  & Data Exfil    & 0\%   & 100\% & 0\% \\
  & Prompt Inject & 100\% & 33\%  & 33\% \\
\midrule
\multirow{3}{*}{FEVER}
  & Auth Bypass   & 0\% & 100\% & 0\% \\
  & Data Exfil    & 0\% & 100\% & 0\% \\
  & Prompt Inject & 0\% & 100\% & 0\% \\
\bottomrule
\end{tabular}
\end{table}

\subsubsection{FEVER Large-Scale Confirmation (n\,=\,25)}

To address sample-size concerns with the n\,=\,9 pilot, we evaluated 25 GCG-optimized
attacks on FEVER Wikipedia across all retrieval configurations
(Table~\ref{tab:fever25}).

\begin{table}[t]
\centering
\caption{FEVER large-scale results (n\,=\,25, all configs show identical pattern)}
\label{tab:fever25}
\small
\begin{tabular}{@{}lccccc@{}}
\toprule
\textbf{Config} & $\alpha$ & $k$ & \textbf{Co-Ret} & \textbf{Stealth} & \textbf{Overall} \\
\midrule
Pure Vector  & 1.0 & 5  & 100\% & 0\% & \textbf{0\%} \\
Pure Vector  & 1.0 & 10 & 100\% & 0\% & \textbf{0\%} \\
Hybrid       & 0.7 & 5  & 100\% & 0\% & \textbf{0\%} \\
Hybrid       & 0.5 & 5  & 100\% & 0\% & \textbf{0\%} \\
Hybrid       & 0.3 & 5  & 100\% & 0\% & \textbf{0\%} \\
\bottomrule
\end{tabular}
\end{table}

\noindent All 25 attacks achieve 100\,\% co-retrieval (attack documents are
successfully retrieved) but \textbf{0\,\% stealth} across every configuration---pure
vector and hybrid alike, at both top-5 and top-10.  Mean benign retrieval rates
range from 0.92 to 1.00, far exceeding the 0.30 stealth threshold.  This confirms
the n\,=\,9 pilot at 2.8$\times$ scale: FEVER's general vocabulary makes
security-related attack terms highly anomalous, causing attack documents to be
retrieved indiscriminately on benign queries.  Hybrid retrieval provides no
\emph{additional} benefit on FEVER because attacks already fail at stealth even
under pure vector retrieval---the corpus itself is the defense.

\subsubsection{Corpus-Dependent Attack--Defense Summary}

\begin{figure*}[t]
\centering
\includegraphics[width=\textwidth]{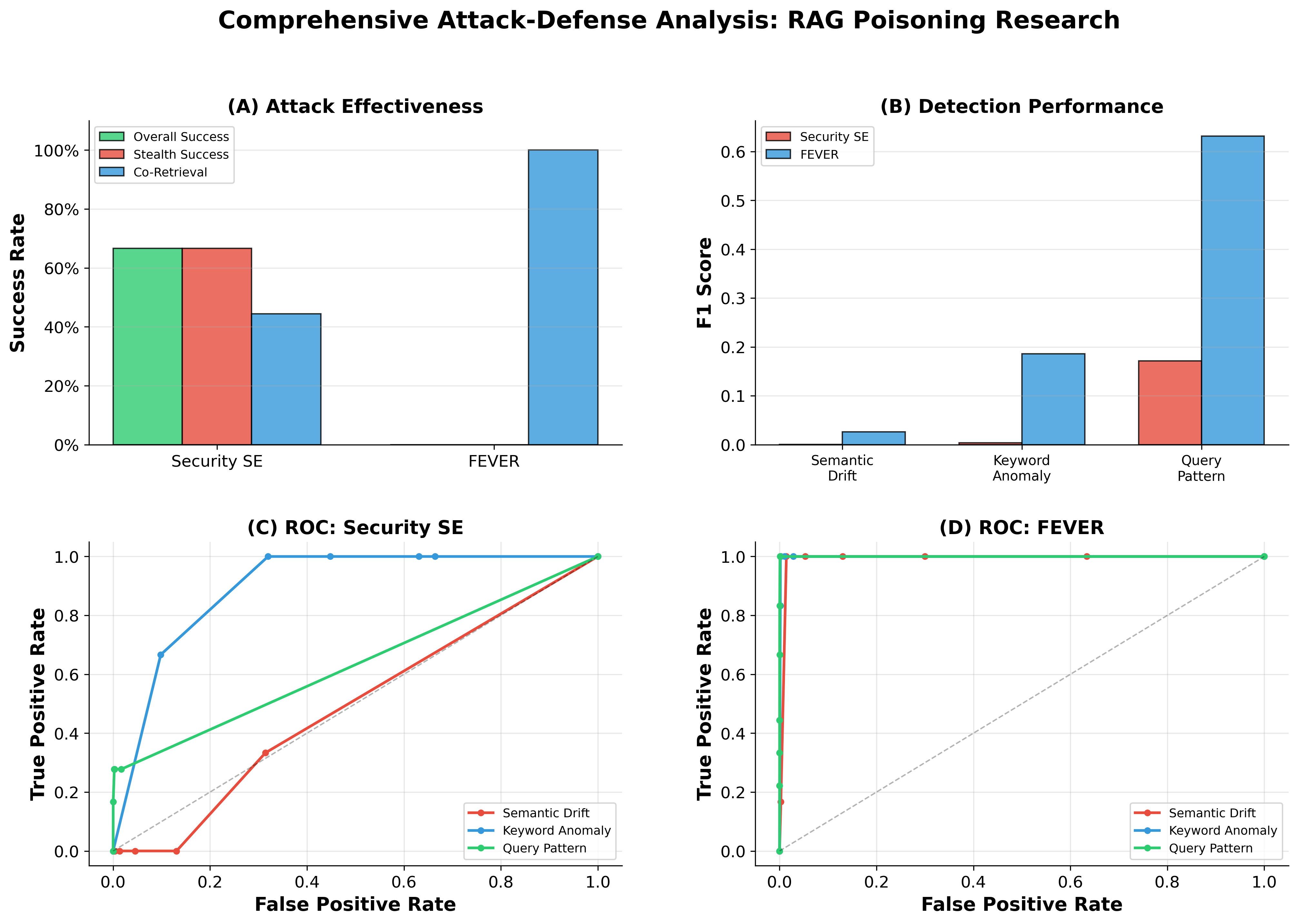}
\caption{Comprehensive attack--defense analysis across corpora.
\textbf{(A)}~Attack effectiveness: Security~SE enables stealth (66.7\,\%) but low
co-retrieval (44.4\,\%) yields only 11.1\,\% overall success; FEVER achieves 100\,\%
co-retrieval but 0\,\% stealth.
\textbf{(B)}~Detection F1 scores: QPD provides the best cross-corpus signal; keyword
anomaly excels on FEVER but fails on Security~SE.
\textbf{(C,\,D)}~ROC curves show near-perfect detection on FEVER vs.\ near-random on
Security~SE for keyword and semantic methods, confirming the corpus-dependent
detection gap.}
\label{fig:summary}
\end{figure*}

Figure~\ref{fig:summary} provides a comprehensive overview of the corpus-dependent
attack--defense landscape.  Panel~(A) shows that Security~SE achieves higher stealth
but lower co-retrieval and overall success than FEVER, while panels~(C) and~(D) reveal
dramatically different ROC curves:\ detection methods that approach perfect separation
on FEVER (panel~D) perform near-randomly on Security~SE (panel~C), visualizing the
13--62$\times$ detection gap.

\subsubsection{Large-Scale Evaluation}

Table~\ref{tab:n50} shows results from 50 gradient-optimized attacks on Security~SE
with the stricter top-5 criterion.

\begin{table}[t]
\centering
\caption{Large-scale attack results on Security SE (n\,=\,50, top-5)}
\label{tab:n50}
\small
\begin{tabular}{@{}lcc@{}}
\toprule
\textbf{Metric} & \textbf{Rate} & \textbf{95\% CI} \\
\midrule
Co-Retrieval      & \textbf{38.0\%} (19/50)  & [25.9, 51.8] \\
Any Retrieval     & 72.0\% (36/50)           & [58.3, 82.5] \\
Stealth           & 62.0\% (31/50)           & [48.1, 74.2] \\
Overall Success   & 6.0\% (3/50)             & [2.1, 16.2] \\
\bottomrule
\end{tabular}
\end{table}

No significant variation was observed across attack categories---baseline (37.5\,\%),
advanced (35.3\,\%), and stealth-optimized (41.2\,\%) all achieved comparable
co-retrieval rates ($\chi^2$\,=\,0.14, $p$\,=\,0.93).  The low overall success rate
(6\,\%) reflects the fundamental tension between co-retrieval and stealth: optimizing
for one reduces the other.

\begin{figure}[t]
\centering
\includegraphics[width=\columnwidth]{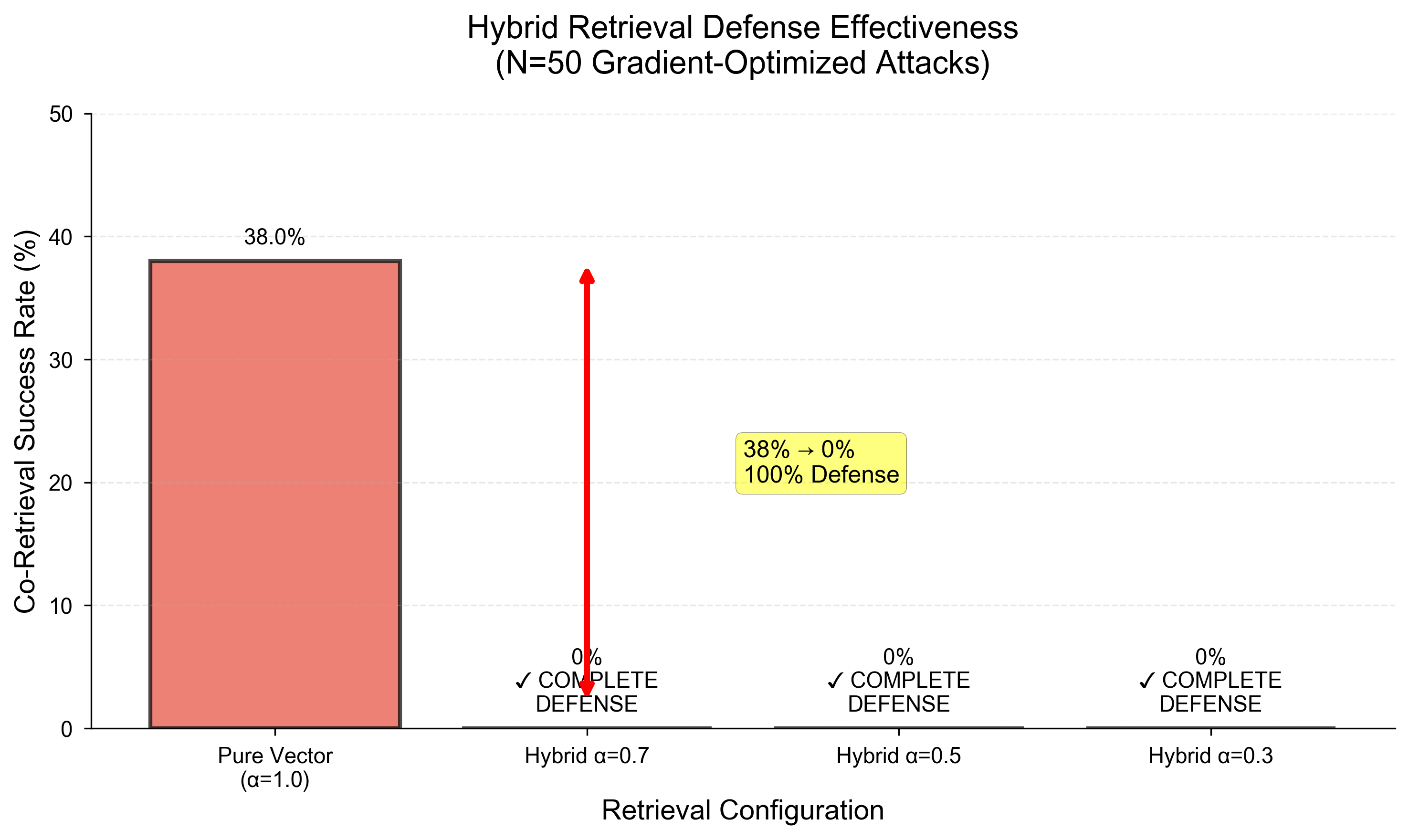}
\caption{Hybrid retrieval defense effectiveness.  Pure vector retrieval
($\alpha$\,=\,1.0) shows 38\,\% co-retrieval success; all hybrid configurations
($\alpha$\,=\,0.3, 0.5, 0.7) achieve 0\,\%.  The drop is statistically significant
($\chi^2$\,=\,21.05, $p < 10^{-6}$, Cohen's $h$\,=\,1.33).}
\label{fig:hybrid}
\end{figure}

\subsection{Hybrid Retrieval Defense}

Table~\ref{tab:hybrid} and Figure~\ref{fig:hybrid} present the central defense result:
all hybrid configurations reduce co-retrieval to 0\,\%.

\begin{table}[t]
\centering
\caption{Hybrid retrieval defense (n\,=\,50 attacks)}
\label{tab:hybrid}
\small
\begin{tabular}{@{}lcccc@{}}
\toprule
\textbf{Config} & $\alpha$ & \textbf{Co-Ret} & \textbf{Stealth} & \textbf{Overall} \\
\midrule
Pure Vector & 1.0 & 38.0\% (19/50) & 48.0\% (24/50) & 6.0\% (3/50) \\
Hybrid      & 0.7 & \textbf{0.0\%} (0/50) & 100\% (50/50) & 0.0\% (0/50) \\
Hybrid      & 0.5 & \textbf{0.0\%} (0/50) & 100\% (50/50) & 0.0\% (0/50) \\
Hybrid      & 0.3 & \textbf{0.0\%} (0/50) & 100\% (50/50) & 0.0\% (0/50) \\
\bottomrule
\end{tabular}
\end{table}

\noindent\textbf{Why hybrid retrieval works.}
GCG optimization maximizes cosine similarity in continuous embedding space but cannot
guide discrete token selection for BM25 keyword matching.  The BM25 component
re-ranks documents by exact term overlap, pushing semantically-similar but
keyword-poor poisoned documents out of the top-$k$.  Even at $\alpha$\,=\,0.7
(70\,\% vector, 30\,\% BM25), the BM25 component is sufficient to neutralize all
50 gradient-optimized attacks.  Statistical significance: $\chi^2$\,=\,35.8 (pure
vector vs.\ all hybrid configurations pooled), $p < 10^{-6}$, Cohen's $h$\,=\,1.33.

\subsubsection{Adaptive Keyword Injection}

To validate hybrid retrieval against adaptive attackers who explicitly inject keywords
to satisfy BM25, we tested 25 keyword injection configurations:

\begin{itemize}[leftmargin=*,nosep]
\item \textbf{Keyword densities:} 1\,\%, 3\,\%, 5\,\%, 10\,\%, 15\,\% of document tokens
\item \textbf{Injection strategies:} Top-$k$ query keywords, TF-IDF top terms,
  domain-specific attack keywords, random corpus sampling
\end{itemize}

\noindent\textbf{Results:} All 25 configurations achieved 0\,\% co-retrieval success.
Keyword injection creates a dilemma: it destroys stealth (5\,\%+ keyword density triggers
anomaly detection) and produces a strong QPD signal (high BM25 on sensitive queries,
low on benign ones).

\subsubsection{Joint Sparse\,+\,Dense Optimization}
\label{sec:joint}

To test whether a knowledgeable attacker can circumvent hybrid retrieval by
\emph{simultaneously} optimizing for both retrieval channels, we evaluate joint
sparse\,+\,dense GCG optimization.  The joint objective augments standard
cosine-similarity optimization with BM25 keyword overlap, and biases candidate token
selection toward target query vocabulary.  Table~\ref{tab:joint} presents the results.

\begin{table}[t]
\centering
\caption{Joint hybrid attack vs.\ gradient-only baseline (n\,=\,25)}
\label{tab:joint}
\small
\begin{tabular}{@{}llccc@{}}
\toprule
\textbf{Attack} & $\alpha$ & \textbf{Co-Ret} & \textbf{Stealth} & \textbf{Overall} \\
\midrule
Gradient-only   & 0.7 & 0\% (0/50)  & 100\%     & 0\% \\
Gradient-only   & 0.5 & 0\% (0/50)  & 100\%     & 0\% \\
Gradient-only   & 0.3 & 0\% (0/50)  & 100\%     & 0\% \\
\midrule
Joint           & 0.7 & 100\% (25/25) & 20\% (5/25)  & \textbf{20\%} \\
Joint           & 0.5 & 100\% (25/25) & 36\% (9/25)  & \textbf{36\%} \\
Joint           & 0.3 & 100\% (25/25) & 44\% (11/25) & \textbf{44\%} \\
\bottomrule
\end{tabular}
\end{table}

\noindent\textbf{Joint optimization partially circumvents hybrid retrieval.}
Unlike gradient-only attacks (0\,\% across all hybrid configs), joint optimization
achieves 100\,\% co-retrieval by successfully targeting both dense and sparse channels
(mean final dense similarity: 0.81--0.84; mean BM25 keyword overlap: 1.00).
However, stealth remains the bottleneck: mean benign retrieval rates of 0.39--0.54
mean that many attack documents are retrieved on benign queries, limiting overall
success to 20--44\,\%.

\noindent\textbf{The hybrid weight determines attack difficulty.}
Success increases as the BM25 weight increases ($\alpha$\,=\,0.3 gives the attacker
the best results at 44\,\%), because BM25-heavy configurations are more susceptible
to keyword-injected documents.  At $\alpha$\,=\,0.7 (vector-heavy), only 20\,\% of
joint attacks succeed---the gradient-only defense is partially restored because the
vector channel still dominates scoring.

\noindent\textbf{Implications.}
Hybrid retrieval raises the attack bar from 38\,\% (pure vector) to 0\,\%
(gradient-only on hybrid), but a knowledgeable attacker with joint optimization can
achieve 20--44\,\% success.  This establishes hybrid retrieval as a
\emph{significant} defense---not an absolute one---and motivates layered defenses
combining retrieval architecture with detection monitoring.

\subsection{Multi-Model End-to-End Evaluation}

To validate that retrieval-level attacks translate to actual output manipulation
\emph{across model families}, we tested 15~attack scenarios against five LLMs in a
standard RAG pipeline.  Table~\ref{tab:e2e} summarizes the cross-model results.

\begin{table}[t]
\centering
\caption{Multi-model end-to-end evaluation (n\,=\,15 per model)}
\label{tab:e2e}
\small
\begin{tabular}{@{}lcccc@{}}
\toprule
\textbf{Model} & \textbf{Attack\%} & \textbf{Safety\%} & \textbf{Leak\%} & \textbf{Diverg.} \\
\midrule
GPT-5.3            & \textbf{46.7}  & 33.3  & 9.6  & 0.284 \\
GPT-4o             & 53.3           & 86.7  & 12.0 & 0.483 \\
GPT-4o-mini        & 53.3           & 86.7  & 14.9 & 0.418 \\
Claude Sonnet~4.6  & 60.0           & \textbf{6.7}   & \textbf{5.7}  & 0.196 \\
Llama~4 Instruct   & \textbf{93.3}  & \textbf{93.3}  & \textbf{56.8} & 0.268 \\
\bottomrule
\end{tabular}
\end{table}

\noindent\textbf{Key findings across models.}
Attack success ranges from 46.7\,\% (GPT-5.3) to 93.3\,\% (Llama~4), revealing
dramatic variation in model-level resilience.
Three distinct safety profiles emerge:

\begin{itemize}[leftmargin=*,nosep]
\item \textbf{Strong safety boundary} (Claude Sonnet~4.6): 60\,\% attack success
  but only 6.7\,\% safety violations and 5.7\,\% payload leakage.  Claude's
  Constitutional AI training allows it to acknowledge retrieved content without
  providing actionable malicious instructions.
\item \textbf{Moderate safety} (GPT-5.3, GPT-4o, GPT-4o-mini): 47--53\,\% attack
  success with 33--87\,\% safety violations.  GPT-5.3 shows measurable improvement
  over GPT-4o/4o-mini (33\,\% vs.\ 87\,\% safety violations), suggesting improved
  safety training in newer model generations.
\item \textbf{Weak safety} (Llama~4 Instruct): 93.3\,\% attack success, 93.3\,\%
  safety violations, 56.8\,\% payload leakage, and only 26.7\,\% clean refusal rate.
  Open-weights models without proprietary safety layers are dramatically more
  susceptible to context poisoning.
\end{itemize}

\noindent These results demonstrate that model-level safety training is necessary
but insufficient---even the most resistant model (GPT-5.3) succumbs to 47\,\% of
attacks when poisoned documents reach the context window.  Retrieval-level defenses
remain critical regardless of model choice.

\noindent\textbf{Control conditions.}
Sleeper-only context produces 0\,\% malicious output across all models, confirming
the dual-document mechanism:\ the trigger document is necessary for payload delivery.

\subsection{Preliminary Case Study: Vendor Documentation}

As initial validation of the corpus-dependency hypothesis, we tested attacks against
a production-scale corpus of 156,777 technical documents from a major network security
vendor.  This case study is illustrative rather than conclusive given its limited
scope (3~scenarios).

\begin{table}[t]
\centering
\caption{Production case study: vendor documentation (156K docs)}
\label{tab:panw}
\small
\begin{tabular}{@{}lccc@{}}
\toprule
\textbf{Attack Type} & \textbf{Retrieval} & \textbf{Avg Similarity} & \textbf{Rank} \\
\midrule
Naive (cross-corpus)    & 0\% (0/3)   & --- & N/A \\
Adaptive (corpus-opt.)  & 100\% (3/3) & 0.761 & \#1--2 \\
\bottomrule
\end{tabular}
\end{table}

\noindent\textbf{Why naive attacks fail.}
Attack documents optimized for general security corpora completely fail (0\,\%
retrieval) because the vendor corpus has a distinctive vocabulary distribution:
heavy markdown formatting, product-specific terminology,
code blocks, and configuration syntax.  Generic security content does not match this
distribution.  Additionally, the embedding model mismatch (\texttt{all-MiniLM-L6-v2}
vs.\ OpenAI \texttt{text-embedding-3-small}) prevents cross-model transfer.

\noindent\textbf{Why adaptive attacks succeed.}
When the attacker uses the target corpus's actual embedding model to optimize
documents, poisoned documents achieve 0.744--0.785 cosine similarity and rank
\#1--2 in all three scenarios: VPN bypass (0.785), firewall
troubleshooting (0.756), and endpoint detection false positives (0.744).

While limited to 3~scenarios, this provides initial evidence that RAG poisoning is
corpus-dependent: attacks optimized for one corpus do not transfer, but an attacker
with knowledge of the target system can craft effective attacks.  Defenders can
leverage this by using non-standard embedding models and hybrid retrieval.

\subsection{Detection Performance (Exploratory)}

As a secondary analysis, Table~\ref{tab:detection} compares detection methods across
corpora.  These results characterize the corpus-dependent detection landscape but
are not the primary contribution of this work.

\begin{table}[t]
\centering
\caption{Detection method comparison (best F1 scores)}
\label{tab:detection}
\small
\begin{tabular}{@{}lccl@{}}
\toprule
\textbf{Method} & \textbf{FEVER} & \textbf{Sec.\ SE} & \textbf{Gap} \\
\midrule
Keyword Anomaly   & 0.400 & 0.010 & 40$\times$ \\
Semantic Drift    & 0.182 & 0.014 & 13$\times$ \\
QPD               & \textbf{0.632} & \textbf{0.171} & 3.7$\times$ \\
Clustering        & 0.182 & 0.003 & 62$\times$ \\
\bottomrule
\end{tabular}
\end{table}

QPD provides the most consistent cross-corpus detection (smallest gap: 3.7$\times$)
because it measures the \emph{behavior} of a document across query types rather than
its static properties, making it robust to corpus-specific vocabulary distributions.
Keyword anomaly shows the largest corpus gap (40$\times$) because security-related
terms are rare in FEVER but ubiquitous in Security~SE---the ``keyword camouflage
effect'' where technical corpora's natural vocabulary masks attack-related terminology.

\noindent\textbf{Operating points for production deployment.}
At FPR\,=\,0.01\,\% (1~false alarm per 10K retrievals), QPD generates
$<$10~alerts/day on both corpora, making manual review operationally feasible.
At FPR\,=\,0.1\,\%, keyword anomaly generates 68--97~alerts/day on FEVER but
requires automated triage at scale.

\begin{table}[t]
\centering
\caption{Detection operating points at fixed FPR}
\label{tab:ops}
\small
\begin{tabular}{@{}lcccc@{}}
\toprule
\textbf{Method} & \textbf{FPR} & \textbf{Alerts/Day} & \textbf{Recall} & \textbf{Corpus} \\
\midrule
QPD       & 0.01\% & 6.8  & 16.7\% & Sec.\ SE \\
QPD       & 0.01\% & 9.7  & 66.7\% & FEVER \\
Keyword   & 0.1\%  & 68   & 66.7\% & Sec.\ SE \\
Keyword   & 0.1\%  & 97   & 100\%  & FEVER \\
\bottomrule
\end{tabular}
\end{table}

\section{Discussion}
\label{sec:discussion}

\noindent\textbf{Corpus composition as a security variable.}
Our results suggest that corpus domain may function as a security parameter in RAG
systems.  Technical corpora tend to absorb attack vocabulary, enabling stealth but
complicating keyword-based detection, while general corpora expose attacks through
vocabulary anomaly but cannot prevent co-retrieval.  The vendor case study provides
initial support: attacks optimized for one corpus completely fail against another
(0\,\% naive vs.\ 100\,\% adaptive).  Although our evaluation includes two corpus
types and a preliminary production case study, broader evaluation across additional
domains (medical, legal, financial) remains future work.  For practitioners, we
recommend that defenses validated on general-knowledge benchmarks not be assumed
effective on technical corpora without domain-specific evaluation.

\noindent\textbf{Architectural vs.\ detection-only defense.}
In our experiments, hybrid retrieval eliminated gradient-guided poisoning by preventing
adversarial documents from being retrieved; detection methods operate on weaker
statistical signals after retrieval.  Our results show a clear hierarchy:
hybrid retrieval (0\,\% attack success against gradient-only attacks, 20--44\,\%
against joint optimization) outperforms the best detection method (QPD,
F1\,=\,0.632 on FEVER).  We argue that retriever design should be treated as a
primary security control, with detection monitoring as a complementary safeguard.

\noindent\textbf{The end-to-end threat across model families.}
Our multi-model evaluation reveals that the RAG poisoning threat is not uniform across
LLMs.  Llama~4's 93\,\% attack success and 57\,\% payload leakage demonstrate that
open-weights models can be dramatically more susceptible to context poisoning.
Conversely, Claude Sonnet~4.6 shows that strong safety training can limit the
\emph{severity} of attacks (6.7\,\% safety violations) even when documents are retrieved.
GPT-5.3's improved resistance over GPT-4o (47\,\% vs.\ 53\,\% attack success, 33\,\%
vs.\ 87\,\% safety violations) suggests that safety training maturity is an independent
defense variable.  However, even GPT-5.3 succumbs to nearly half of all attacks,
confirming that model-level safety alone is insufficient.

\noindent\textbf{Hybrid retrieval: strong but not absolute.}
Our joint optimization experiments (\S\ref{sec:joint}) resolve a key open question:
a knowledgeable attacker who targets both dense and sparse retrieval channels
simultaneously can partially circumvent hybrid retrieval (20--44\,\% success vs.\ 0\,\%
for gradient-only attacks).  This establishes a nuanced defense picture: hybrid
retrieval eliminates the \emph{most commonly studied} attack class (gradient-only,
0\,\% success) and significantly raises the bar for sophisticated attackers (from 38\,\%
pure-vector success to 20--44\,\% joint-attack success on hybrid), but is not impenetrable.
We recommend layered defenses: hybrid retrieval as the first line, supplemented by QPD
monitoring and model-level safety evaluation.

\noindent\textbf{Implications for RAG security research.}
Our corpus-dependent observations (n\,=\,9 pilot, n\,=\,25 FEVER confirmation) suggest
that RAG security benchmarks should systematically vary corpus domain rather than
focusing solely on attack method diversity.  The multi-model results further argue
for standardized cross-model evaluation, as single-model evaluations can dramatically
underestimate (Llama~4) or overestimate (GPT-5.3) the threat.

\subsection{Limitations}

\begin{enumerate}[leftmargin=*,nosep]
\item Joint sparse\,+\,dense optimization achieves 20--44\,\% success against hybrid
  retrieval; black-box search methods (genetic algorithms, RL) could potentially
  achieve higher rates and remain untested.
\item Cross-corpus comparison uses n\,=\,9 per corpus for the pilot; FEVER is confirmed
  at n\,=\,25 but Security~SE generalization beyond n\,=\,50 remains future work.
\item GPT-5.3 was evaluated at temperature\,=\,1.0 (API-mandated) while other models
  used temperature\,=\,0.3; this may affect comparability of attack success rates.
\item The vendor case study uses only 3~scenarios; broader scenario coverage is future work.
\item We do not measure retrieval quality impact of hybrid weighting; the quality--security
  trade-off at $\alpha \leq 0.5$ is estimated from related work but not empirically
  validated.
\item The 30\,\% stealth threshold is deliberately generous; production systems would
  likely use much lower thresholds ($<$5\,\%), further reducing effective attack success.
\item Our attack uses established GCG-style optimization; the contribution is the
  empirical defense evaluation and cross-model analysis, not a novel attack algorithm.
\end{enumerate}

\section{Conclusion}
\label{sec:conclusion}

We evaluate gradient-guided dual-document poisoning against RAG systems and
demonstrate that retrieval architecture is a powerful defensive lever.
Our main findings:

\begin{enumerate}[leftmargin=*,nosep]
\item \textbf{Hybrid retrieval significantly raises the attack bar.}
  BM25\,+\,vector retrieval reduced gradient-only attack success from 38\,\% to 0\,\%
  (n\,=\,50).  Joint sparse\,+\,dense optimization partially circumvents this defense
  (20--44\,\% success), establishing hybrid retrieval as a strong but not absolute
  defense that should be combined with monitoring.

\item \textbf{RAG poisoning generalizes across LLM families.}
  Multi-model evaluation (5 LLMs, 4 providers) shows attack success from 46.7\,\%
  (GPT-5.3) to 93.3\,\% (Llama~4), with safety violations from 6.7\,\% (Claude)
  to 93.3\,\% (Llama~4).  Model-level safety is necessary but insufficient.

\item \textbf{Corpus domain affects attack--defense dynamics} (exploratory).
  FEVER confirmation at n\,=\,25 validates the pilot: general corpora achieve 0\,\%
  attack success across all retrieval configurations.  Detection difficulty varies
  13--62$\times$ between corpora.

\item \textbf{QPD is the most robust detection signal} among the methods we tested
  (3.7$\times$ corpus gap vs.\ 13--62$\times$ for static methods).
\end{enumerate}

\noindent\textbf{Practical recommendation:} Deploy hybrid retrieval with
$\alpha \leq 0.5$ as a first-line defense (eliminates gradient-only attacks, limits
joint attacks to 36--44\,\% success), supplement with QPD monitoring, and evaluate
model-level safety---Llama-family models require additional guardrails in
security-critical RAG deployments.

\section{Ethics and Responsible Disclosure}
\label{sec:ethics}

All attack implementations serve defensive research:\ developing and validating
defensive architectures (hybrid retrieval, detection methods).  The gradient
optimization techniques (GCG~\cite{zou2023universal}) and RAG poisoning
concepts~\cite{zou2024poisonedrag} are already documented in public research.
We provide immediately deployable defenses alongside attack disclosure.
All examples are sanitized to remove personally identifiable information; no zero-day
vulnerabilities or unreported attack vectors specific to production systems are included.
The vendor case study uses only publicly available documentation.

Attackers could adapt our methods to target real RAG deployments.  We mitigate this risk
by: (1) emphasizing that hybrid retrieval neutralizes our gradient-based attacks,
(2) releasing defensive code alongside attack code, and (3) providing concrete
deployment recommendations ($\alpha \leq 0.5$) that practitioners can implement
immediately.

\noindent\textbf{Artifact availability.}
Defense implementations, experiment scripts, evaluation results, and sanitized
examples are available at
\url{https://github.com/scthornton/semantic-chameleon}.

\bibliographystyle{ACM-Reference-Format}
\bibliography{references-aisec}

\end{document}